\documentclass[journal]{IEEEtran}

\usepackage[english]{babel}
\usepackage[utf8]{inputenc}
\usepackage[T1]{fontenc}

\usepackage{amsmath}
\usepackage{graphicx}
\usepackage[colorinlistoftodos]{todonotes}
\usepackage[colorlinks=true, allcolors=blue]{hyperref}
\usepackage{float}
\usepackage{authblk}

\let\vec\mathbf
\newcommand{\figWidth}{\linewidth}

\title{Error Correction in Structured Optical Receivers}
\author[1]{Alec M. Hammond}
\author[2]{Ian W.  Frank}
\author[1]{Ryan M. Camacho \thanks{Corresponding Author: camacho@byu.edu}}
\affil[1]{Department of Electrical and Computer Engineering, Brigham Young University, Provo UT 84602} 
\affil[2]{Draper Laboratory, Cambridge, MA 02139}
\begin{document}


\maketitle


\begin{abstract}
	 Integrated optics Green Machines enable better communication in photon-starved environments, but fabrication inconsistencies induce unpredictable internal phase errors, making them difficult to construct. We describe and experimentally demonstrate a new method to compensate for arbitrary phase errors by deriving a convex error space and implementing an algorithm to learn a unique codebook of codewords corresponding to each matrix.
\end{abstract}{}
\section{Introduction}
Recent developments in structured optical receivers demonstrate potential to exceed classical information transmission limits in photon-starved environments \cite{guha_structured_2011}. Rather than reading codewords symbol by symbol, these networks perform joint-detection over multiple symbols simultaneously. Consequently, the Shannon capacity of this device approaches the Holevo limit --- an upper bound for communication transceivers relying on quantum state alphabets.\par 

One particular architecture, known as the "Green Machine" (GM) in the literature, matches the topology of the fast Fourier transform (FFT) by mapping the inputs and outputs of various beamsplitter stages in a butterfly fashion. It performs a Hadamard transform on a binary phase-shift keyed (BPSK) codebook to generate a pulse-position modulation (PPM) based codebook. So long as we choose a BPSK codebook with columns of a Hadamard matrix as its codewords, the GM architecture will efficiently route each codeword to a single output channel, enabling deep space applications with strict signal power constraints. \par

Leveraging years of process refinement within the semiconductor industry allows silicon photonics to provide an ideal platform for fabricating the GM \cite{chrostowski_silicon_2015}. These small, solid-state  systems can transition from prototype to large-scale, cost-effective fabrication almost instantly, paving the way for rapid innovation. Constructing this receiver is difficult, however, because preserving the Hadamard transform requires phase matching across the entire network.  

In this paper, we propose and demonstrate a solution to this challenge that allows for complete phase compensation \textit{after} fabrication and applies to both integrated and bulk optical systems.  It is known that in the case of quantum-limited detection, the GM architecture is capable of distinguishing between a maximally orthogonal set of phase-encoded codewords with near ideal performance.  Here we demonstrate that the maximal orthogonality of the codewords is robust to internal phase errors so long as a new codebook is defined that compensates for the internal phase errors. In other words, the GM does not require a strict Hadamard transform for maximum codeword distinction. This is the key observation of this work, and transfers the experimental task to learning the new codewords rather than phase-matching. We present a feedback algorithm that learns the codewords of an arbitrary GM and demonstrate the algorithm's success on a free-space setup.\par

We generalize the GM transformation requirements by relating them to Optical Butler Matrices (OBM). Butler matrices are a class of passive beamforming networks that take a single beam as an input and coherently generate multiple output beams with a well-defined phase relationship \cite{jesse_butler_beam-forming_1961}. First invented over 50 years ago, they are now commonly used at radio frequencies (RF) for applications in beam steering, astronomy, satellite communications, and remote sensing\cite{fenn_development_2000}. These devices are a subclass of the GM architecture, where "phase errors" are strategically placed to induce the proper "codewords", or phase profiles. \par

The outline of the rest of the paper is as follows: In section 2 we give a theoretical overview of the Green Machine and fast Fourier transform. In Section 3 we present our feedback algorithm to find new codewords for a non phase-matched Green Machine and in Section 4 we present our experimental results using the algorithm in a 4 $\times$ 4 system.  Section 5 then ends with concluding remarks and an outlook for future work.

    
\section{Green Machine \& Fast Fourier Transform}

By examining the Green Machine's corresponding transformation matrix and how it matches the FFT topology, we show that any arbitrary device maintains an orthogonal book of codewords. Due to fabrication constraints, we first analyze a GM built with symmetric beamsplitters (i.e. directional couplers). We then extend this analysis to the original Hadamard GM and to Butler Matrices. In the process, we show that any arbitrary device composed of symmetric beamsplitters, regardless of the induced phase errors between stages, has a unique codebook.\par

The original FFT as proposed by Cooley and Tukey is a "divide and conquer" algorithm that recursively simplifies the discrete Fourier transform into more computationally attractive operations. The simplest method, known as radix-2, forms 2-input, 2-output "butterflies" as the base of its recursive routine. These butterflies mix the inputs in a fashion similar to that of a beamsplitter. Each butterfly also applies various  twiddle factors (phase shifts) to the inputs and/or outputs to shape the desired Fourier transform response. These twiddle factors are analogous to the "phase errors" that are present in GM devices after fabrication. \par

Each butterfly can be modeled by a transformation matrix. To find the transformation matrix and butterfly representation of a GM using ideal directional couplers and no phase errors, we recursively cascade matrices that characterize its various components. For example, each two-port directional coupler stage is modeled by
\begin{equation}
A_2 =
\begin{bmatrix}
t & ir \\
ir & t \\
\end{bmatrix}
\end{equation}
where $i$ is $\sqrt{-1}$, $t$ is the Fresnel transmission coefficient, and $r$ is the Fresnel reflection coefficient. We model larger networks by cascading a circulant matrix, $C_n$, with a combined beamsplitter stage:
\begin{equation}
A_n = C_n  (A_{\frac{n}{2}} \bigotimes
\begin{bmatrix}
1 & 0 \\
0 & 1 \\
\end{bmatrix}
)
\end{equation}
so that an ideal 4 port device where $t=r=\frac{\sqrt{2}}{2}$ has the following transformation matrix:
\begin{equation}
\begin{split}
A_4 &=
\frac{1}{2}
\begin{bmatrix}
1 & 0 & i & 0 \\
0 & i & 0 & 1 \\
i & 0 & 1 & 0 \\
0 & 1 & 0 & i \\
\end{bmatrix}
\begin{bmatrix}
1 & i & 0 & 0 \\
i & 1 & 0 & 1 \\
0 & 0 & 1 & i \\
0 & 0 & i & 1 \\
\end{bmatrix}\\
&=
\frac{1}{2}
\begin{bmatrix}
1 & i & i & -1 \\
i & 1 & -1 & i \\
i & -1 & 1 & i \\
-1 & i & i & 1 \\
\end{bmatrix}
\end{split}
\end{equation}
A codeword is the list $n$ phases of the coherent inputs to the $n$ channels of the GM that result in all of the beams combined into a single output. For an $n$-channel GM there are $n$ codewords, each one corresponding to combining the beam to a unique output. We extract a device's codebook, $H$ mathematically by inverting its corresponding transfer matrix:
\begin{equation}
\begin{split}
H_4 &=
\left\lbrace
\frac{1}{2}
\begin{bmatrix}
1 & i & i & -1 \\
i & 1 & -1 & i \\
i & -1 & 1 & i \\
-1 & i & i & 1 \\
\end{bmatrix}
\right\rbrace^{-1}
\begin{bmatrix}
2 & 0 & 0 & 0 \\
0 & 2 & 0 & 0 \\
0 & 0 & 2 & 0 \\
0 & 0 & 0 & 2 \\
\end{bmatrix}
\\
&= 
\begin{bmatrix}
1 & -i & -i & -1 \\
-i & 1 & -1 & -i \\
-i & -1 & 1 & -i \\
-1 & -i & -i & 1 \\
\end{bmatrix}
\end{split}
\end{equation}
The cascaded diagonal matrix corresponds to the expected output field magnitude for each codeword. The codewords are then the columns of $H$. In this case, all of the codewords are mutually orthogonal (since the codebook and transformation matrix are unitary) and symbols are separated by $\frac{\pi}{2}$ in phase space.

We can follow a similar procedure to formulate the transformation matrix and corresponding codebook of a Hadamard GM. The device is composed of asymmetric beamsplitter stages, which are each modeled by
\begin{equation}
A_2 =
\frac{\sqrt{2}}{2}
\begin{bmatrix}
1 & 1 \\
1 & -1 \\
\end{bmatrix}
\end{equation}
Upon closer examination, we realize we can build an asymmetric beamsplitter by simply applying phase shifts to the inputs and outputs of the ideal symmetric beamsplitter:
\begin{equation}
\frac{\sqrt{2}}{2}
\begin{bmatrix}
1 & 1 \\
1 & -1 \\
\end{bmatrix}
=
\begin{bmatrix}
-i & 0 \\
0 & -1 \\
\end{bmatrix}
\frac{\sqrt{2}}{2}
\begin{bmatrix}
1 & i \\
i & 1 \\
\end{bmatrix}
\begin{bmatrix}
i & 0 \\
0 & 1 \\
\end{bmatrix}
\end{equation}
Similarly, we model a four-port Hadamard GM by cascading appropriate phasor matrices ($P$) in between stages:
\begin{equation}
A_n = C_n P_n (A_{\frac{n}{2}} \bigotimes
\begin{bmatrix}
1 & 0 \\
0 & 1 \\
\end{bmatrix}
)
\end{equation}
so that resulting transformation matrix is simply a Hadamard matrix:
\begin{equation}
A_4 =
\frac{1}{2}
\begin{bmatrix}
1 & 1 & 1 & 1 \\
1 & -1 & 1 & -1 \\
1 & 1 & -1 & -1 \\
1 & -1 & -1 & 1 \\
\end{bmatrix}
\end{equation}
We determine the corresponding codebook to be
\begin{equation}
H_4 =
\begin{bmatrix}
1 & 1 & 1 & 1 \\
1 & -1 & 1 & -1 \\
1 & 1 & -1 & -1 \\
1 & -1 & -1 & 1 \\
\end{bmatrix}
\end{equation}
As we expect, all codewords are mutually orthogonal and each symbol is separated by $\pi$ in phase space.\par

We model a Butler Matrix the same way we modeled our ideal and Hadamard Green Machines. We note that the fundamental stage also relies on an asymmetric beamsplitter, but the induced phase shifts in between stages are more involved \cite{ueno_systematic_1981}. The final transformation matrix and codebook are given by
\begin{equation}
\frac{\sqrt{2}}{2}
\begin{bmatrix}
1 & i\\
1 & -i\\
\end{bmatrix} =
\begin{bmatrix}
-i & 1\\
0 & -1\\
\end{bmatrix}
\frac{\sqrt{2}}{2}
\begin{bmatrix}
1 & i\\
i & 1\\
\end{bmatrix}
\begin{bmatrix}
i & 0\\
0 & i\\
\end{bmatrix}
\end{equation}
and
\begin{equation}
H_4 =
\begin{bmatrix}
1 & 1 & 1 & 1 \\
1 & i & -1 & -i \\
1 & -1 & 1 & 1 \\
1 & -i & -1 & i \\
\end{bmatrix}
\end{equation}
Here we see that codewords continue to be mutually orthogonal and each symbol is seperated by $\frac{\pi}{2}$ in phase space. 

\begin{figure*}
\centering
\includegraphics[width=0.75\figWidth]{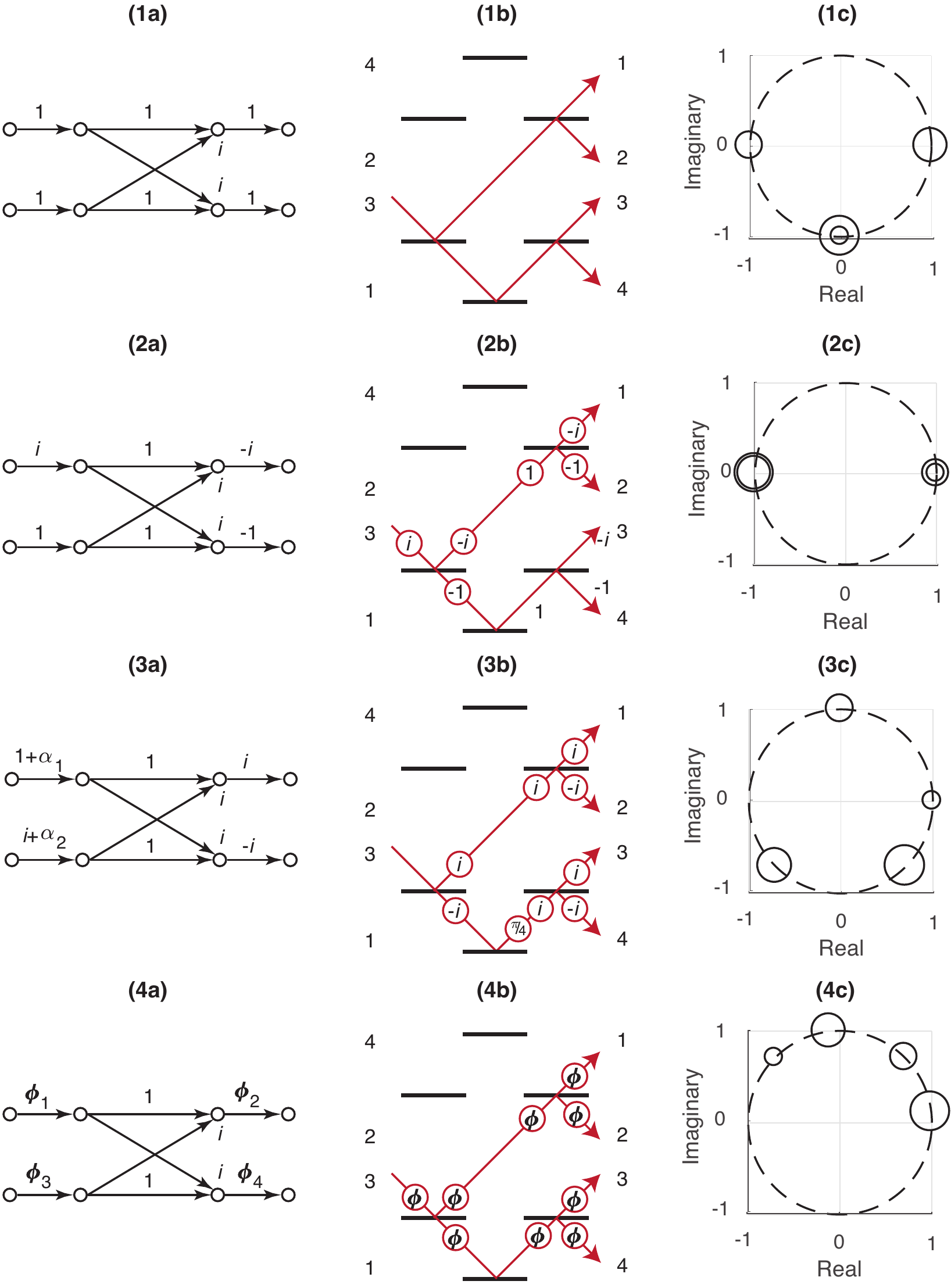}
\caption{A comparison between the equivalent FFT butterfly circuit using directional couplers (column 1), the corresponding GM architecture with effective phase shifts (column 2), and the resulting codewords (column 3) for the ideal GM (row a), the Hadamard GM (row b), the Butler Matrix (row c), and a GM with arbitrary phase errors (row d). The symbol spread for each codeword depends on the embedded phase shifts/errors. For example, the Hadamard GM contains phase errors where all of the symbols are $\pi$ apart, forming BPSK codewords. All other schemes rotate the codewords, but manage to maintain at least $pi/2$ distance between critical stages.}
\label{fig:polarPlots}
\end{figure*}

After exploring all three cases, we realize that the "phase errors" within the device simply rotate the codeword symbols in phase space, but each codeword retains maximum orthogonality. For a more rigorous analysis, we inject arbitrary phase errors between two stages of a 4 port device in Equation \ref{eqn:phaseError} and solve for its accompanying codebook:
\begin{figure*}
\begin{equation}
\label{eqn:phaseError}
\begin{split}
A_4 &= C_4P_4
\begin{bmatrix}
A_2 & 0\\
0 & A_2\\
\end{bmatrix}
\\
&=
\begin{bmatrix}
t & 0 & ir & 0 \\
0 & ir & 0 & t \\
ir & 0 & t & 0 \\
0 & t & 0 & ir \\
\end{bmatrix}
\begin{bmatrix}
e^{i\phi_1} & 0 & 0 & 0 \\
0 & e^{i\phi_2} & 0 & 0 \\
0 & 0 & e^{i\phi_3} & 0 \\
0 & 0 & 0 & e^{i\phi_4} \\
\end{bmatrix}
\begin{bmatrix}
t & ir & 0 & 0 \\
ir & t & 0 & t \\
0 & 0 & t & ir \\
0 & 0 & ir & t \\
\end{bmatrix}
\end{split}
\end{equation}
\end{figure*}

\begin{equation}
\label{eqn:codebook}
\begin{split}
H &= A_4^{-1}X_0 \\
&= 
\frac{1}{2}\begin{bmatrix}
  e^{i\theta_1} & -ie^{i\theta_1} & -ie^{i\theta_2} & -e^{i\theta_2} \\
-ie^{i\theta_1} & -e^{i\theta_1} & e^{i\theta_2} & -ie^{i\theta_2} \\
-ie^{i\theta_3} &  e^{i\theta_3} & -e^{i\theta_4} & -ie^{i\theta_4} \\
-e^{i\theta_3} & -ie^{i\theta_3} & -ie^{i\theta_4} & e^{i\theta_4} \\
\end{bmatrix} 
\begin{bmatrix}
2 & 0 & 0 & 0\\
0 & 2 & 0 & 0 \\
0 & 0 & 2 & 0 \\
0 & 0 & 0 & 2 \\
\end{bmatrix}\\
&=
\begin{bmatrix}
e^{i\theta_1} & -ie^{i\theta_1} & -ie^{i\theta_2} & -e^{i\theta_2} \\
-ie^{i\theta_1} & -e^{i\theta_1} & e^{i\theta_2} & -ie^{i\theta_2} \\
-ie^{i\theta_3} & e^{i\theta_3} & -e^{i\theta_4} & -ie^{i\theta_4} \\
-e^{i\theta_3} & -ie^{i\theta_3} & -ie^{i\theta_4} & e^{i\theta_4} \\
\end{bmatrix}
\end{split}
\end{equation}
We note that even in the presence of arbitrary phase errors, there still exists a set of codewords that lead to maximally orthogonal outputs. One might suspect that the new codewords have become less distinguishable from those without errors. However, we note that each codeword can always be grouped such that the top and bottom halves of the old codewords get rotated in phase space while maintaining a constant distance in phase space. In other words, thanks to the nature of the GM's beamsplitters, phase errors simply shift the device's transfer function from one stable state to another stable state, guaranteeing a new unique set of codewords. Figure \ref{fig:polarPlots} illustrates the equivalent butterfly circuits for each of the above cases, along with the corresponding beamsplitter topology and codeword after exciting the third input port.

We generalize this conclusion to any size Butler Matrix by examining its transformation matrix composition. Since each component of the Butler Matrix, including the phase errors, is modeled by a unitary matrix, the subsequent transformation matrix must also be unitary. The inverse of that transformation matrix (i.e. the codebook) must have orthonormal columns, indicating that each codeword is indeed maximally orthogonal. \par

As an added benefit, we can also model amplitude errors between the stages and find pairings in the elements comprising the codewords such that the losses can also be compensated. Imperfect splitting ratios, for example, are modeled with the $t$ and $r$ coefficients and simply scale the codewords away from unit amplitude. The resulting transformation matrix is composed of orthogonal columns, but is no longer unitary. While it is possible to scale for such errors by also modulating the amplitude of the input, it is not very practical. It should also be noted that amplitude errors, unlike phase errors, result in decreased mutual information and are therefore detrimental to applications for quantum receivers.


\section{Feedback Algorithm}
Experimentally, we would like an algorithm that allows us to inject light into the input ports and find the codewords only by measuring the intensity, but not the phase, at each output. We modulate the phase angle of each input,but not the amplitude, which corresponds to a vector $\vec{x}$ of complex values. These inputs are then modified by the GM, modeled by the transformation matrix $A$. The photodiodes measure the optical power at each output. With these parameters in mind, we designed a feedback loop using a convex objective function and optimization routine.   \par

\begin{figure*}
\centering
\includegraphics[width=0.75\figWidth]{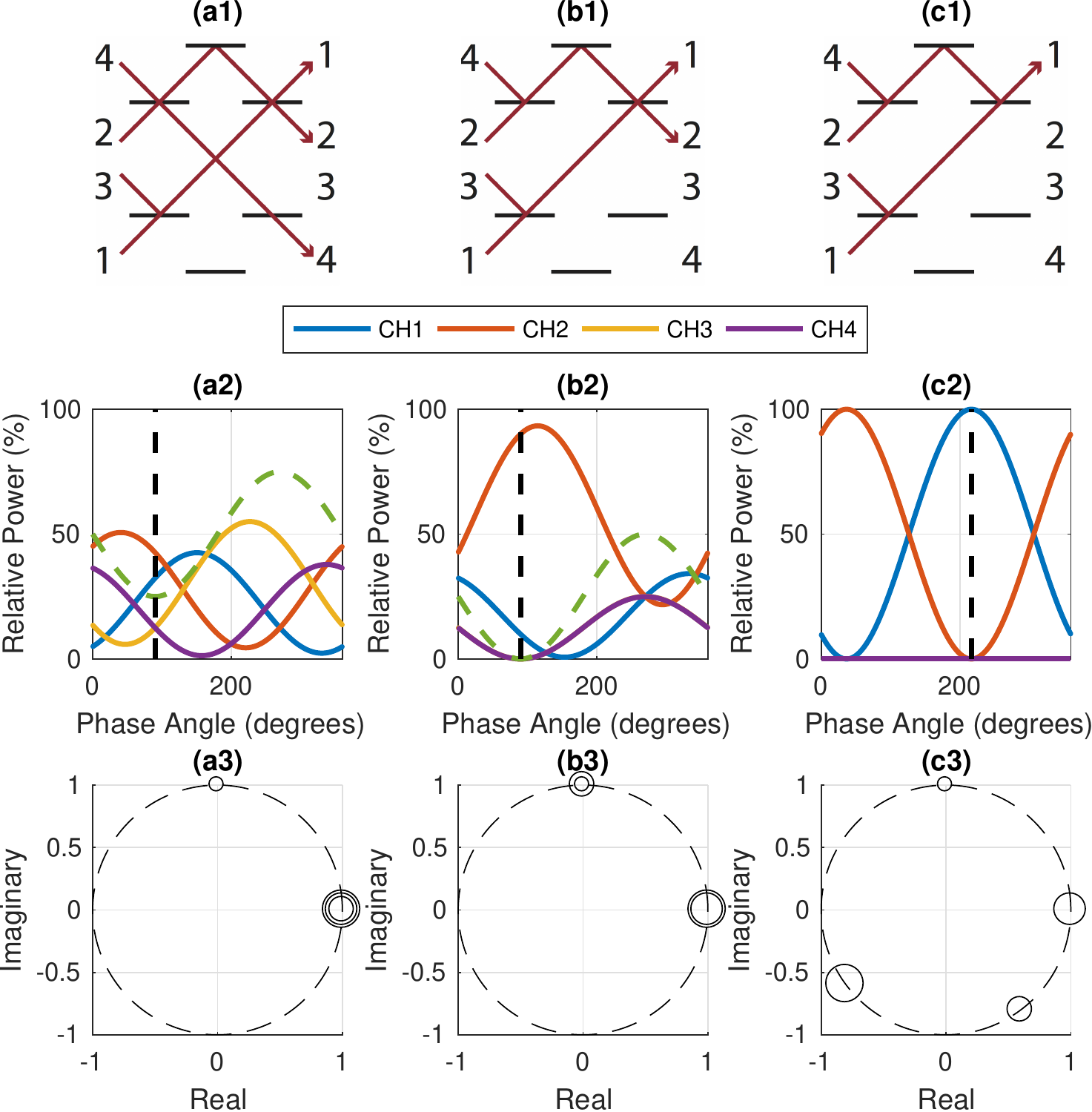}
\caption{A systematic algorithm for determining the first codeword for an arbitrary $4\times4$ GM (i.e. all the light exits port 1). Step 1 is to interfere channels 1 and 3 such that transmission toward ports 3 and 4 is eliminated (a1). This is done by phase modulating channel 1 from $0$ to $2\pi$ (a2) until the combined light of ports 3 and 4 is minimized (shown by dashed green line). The phase that minimizes this quantity is saved and applied to the codeword (a3). The next step is to interfere channels 2 and 4 such that, once again, the transmission towards ports 3 and 4 is minimized (b1). This is similarly done by phase modulating channel 2 until the combined light in ports 3 and 4 is minimized. The corresponding phase is then applied to the codeword (b3). At this point, light only exits ports 1 and 2. We solve for the final codeword by sweeping channels 1 and 3 in concert (to preserve the previously determined necessary phase relationship for interference) until all the light exits port 1. (c1,c2). The resulting phase profile describes the first codeword(c3).}
 \label{fig:simulation}
\end{figure*}

Before we present our convex optimization algorithm, we first present various other phase-learning algorithms to help establish some intuition and perspective on the nature of the system. Consider first an algorithm which leverages the recursive structure of the GM and is shown graphically in \ref{fig:simulation}. By systematically adjusting the input phases so that light at each beamsplitter exits through a single output, we can can find the input phases for a given codeword. Repeating this process for each port produces the codebook for the GM.  While intuitive, this algorithm scales poorly with size and is difficult to implement in hardware. In addition, this algorithm requires prior knowledge of the channel mappings and internal binary combinations within the GM, which can be problematic since several equivalent choices are possible. We therefore seek an algorithm that overcomes these limitations.

A stochastic excitement technique, where each channel is randomly modulated and the statistics of the output are used to solve for the elements of the corresponding A-matrix, would be ideal except for the nonlinearities of the system, which reduces the otherwise convex optimization problem to a non-convex problem that scales poorly with size.  When attempting this approach experimentally, we found the algorithm often did not converge, but became trapped in a local minimum. Likewise, a Maximum Likelihood Estimator (MLE) also reduced to a non-convex optimization problem. Even under ideal conditions, we determined that a both these statistics-based approach are not sufficiently robust. \par  

Fortunately, despite the nonlinearities of the transformation, we can derive a convex objective function, which encourages a brute force optimization approach. We identify the first codeword of a $4\times4$ Butler Matrix, for example, by iterating through various phase profiles that maximize the intensity of output port 1 and minimize the intensity of ports 2, 3, and 4. The output $\vec{y}$ for a given input $\vec{x}$ is given by 
\begin{equation}
A\vec{x} = \vec{y}
\end{equation}
where $\vec{x}$ and $\vec{y}$ are complex. The total intensity measured at all the photodetectors is subsequently modeled as an induced inner product between the output phase profile: 
\begin{equation}
I = \vec{x}^H A^H A \vec{x} 
\end{equation}
The intensity measured at a single photodiode can be modeled with a weighting matrix:
\begin{equation}
I_1 = \vec{x}^H A^H  W_1 A\vec{x}
\end{equation}
If we were interested in channel 1, for example, we would insert the vector $[1,0,0,0]$ along the diagonal of $W$. When optimizing for the second codeword, the second element  along the diagonal of the weighting matrix is 1. So long as we restrict our search space to the complex unit sphere (i.e. we don't modulate the amplitude) the transformation is convex \cite{polyak_convexity_1998}. The optimization problem for the first codeword is described by:
\begin{equation}
\begin{aligned}
& \underset{\vec{x}}{\text{minimize}}
& & J_1(\vec{x}) \\
& \text{subject to}
& & |\vec{x}| = 1 \\
\end{aligned}
\end{equation}
The cost function for the first codeword is then defined as the weighted inner product 
\begin{equation}
J_1(\vec{x}) = -\vec{x}^H A^H  W_1 A\vec{x}
\end{equation}
and the negative sign is necessary to transform the maximization criteria into a minimization problem. \par

\begin{figure}
\centering
\includegraphics[width=\figWidth]{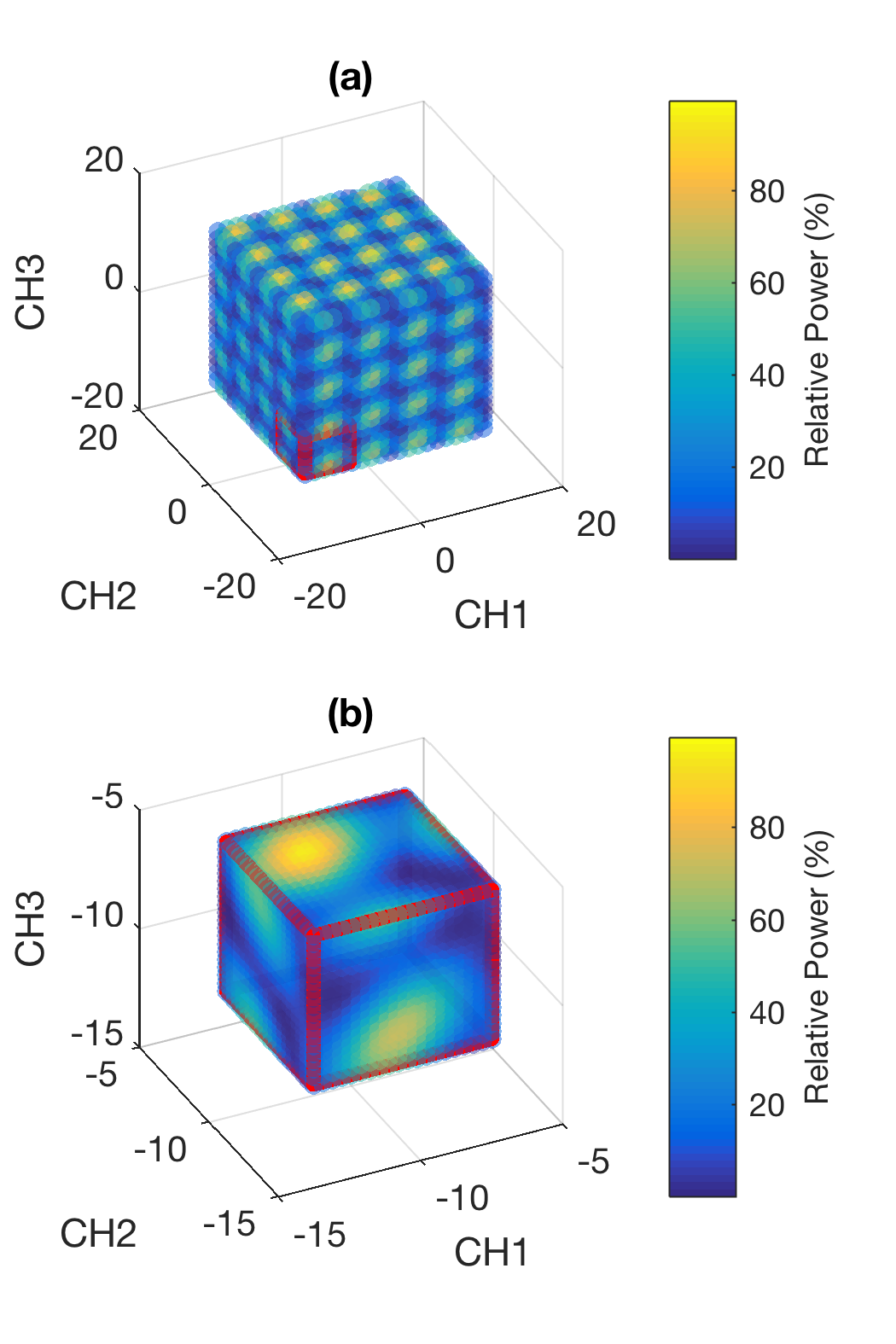}
\caption{(a) The error space for an arbitrary GM with induced phase errors. In this simulation, we phase modulate each channel from $-4\pi$ to $4\pi$. Each "cell" with length of $2\pi$ contains a single valid solution. (b) A single cell where each channel is phase modulated from $-4\pi$ to $-2\pi$.}
%
%

 \label{fig:errorSpace}
\end{figure}

Figure \ref{fig:errorSpace} illustrates the error space for an GM with random phase errors. Since the phase profiles are all relative to each other, there are infinitely many local minima, each of which are valid solutions. Theoretically, the control algorithm can begin descending wherever it initializes and will always converge to a valid solution. Equipment constraints, however, along with environmental noise will limit the phase angle range over which the algorithm can iterate.

\section{Experiment and Results}
To test the control algorithm, we implemented a free-space, four port Butler Matrix using beamsplitters and mirrors arranged in a Michelson-like interferometeric configuration. While the algorithm is intended for eventual use in integrated GM's, our free space implementation allows for a more complete analysis and debugging, since each path can be directly and independently manipulated. \par

\begin{figure}
    \centering
    \includegraphics[width=\figWidth]{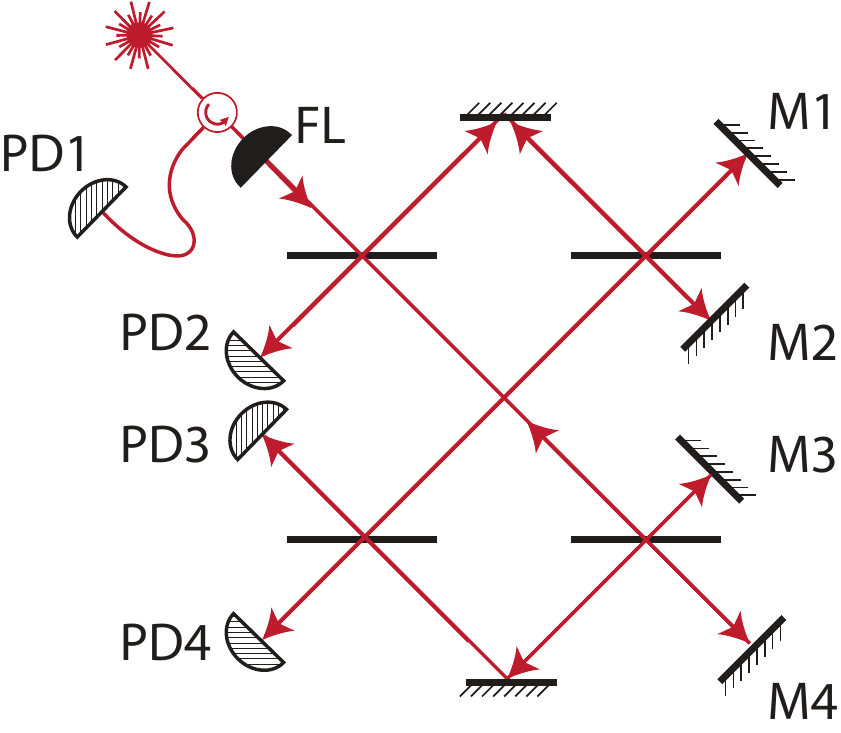}
    \caption{Diagram of the optical paths from the fiber launch (FL) to photodiodes (PD's) in experimental 4 $\times$ 4 BM.}
    \label{fig:setup_a}
    \end{figure}

\begin{figure}
    \centering
    \includegraphics[width=0.9\figWidth]{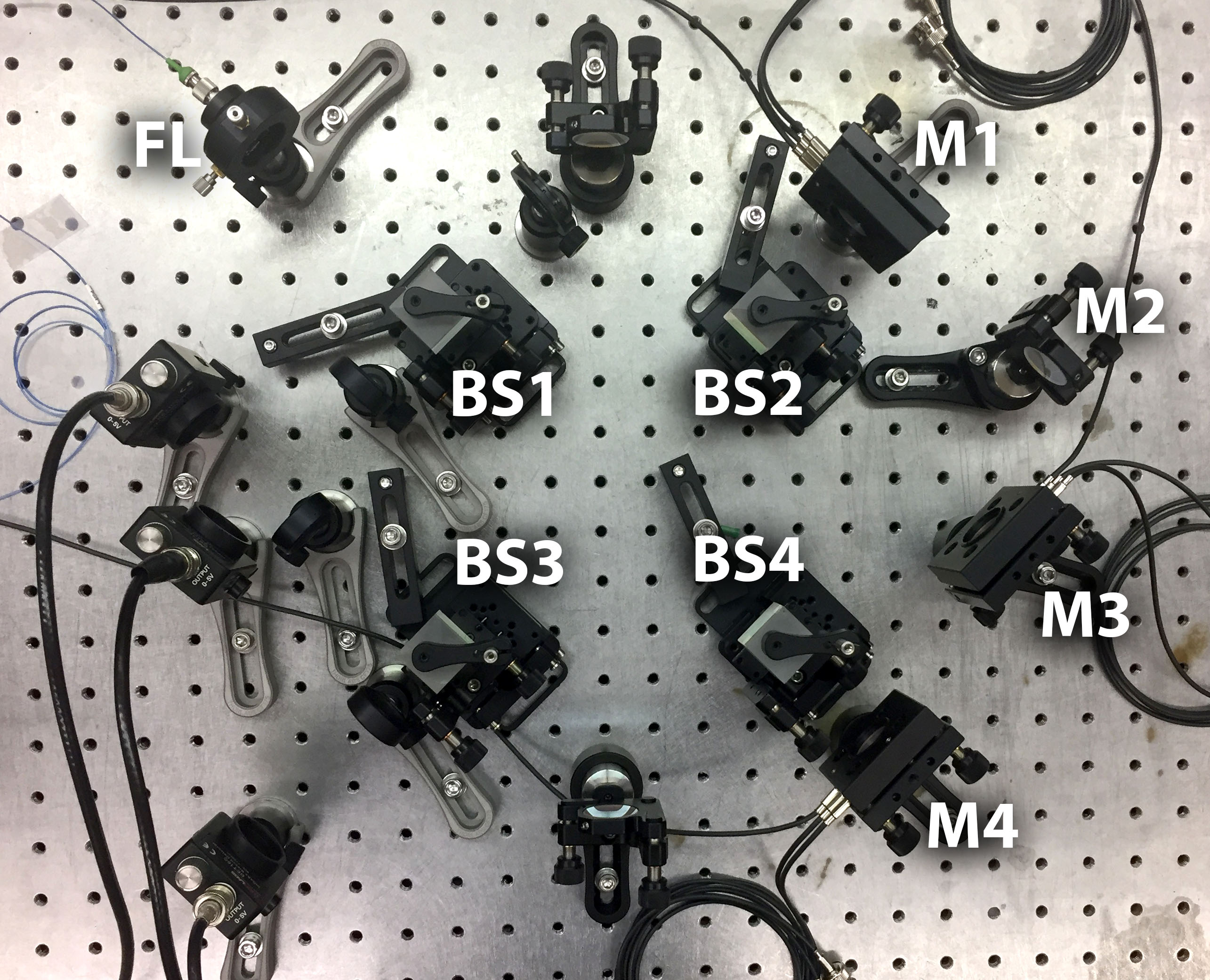}
    \caption{Top view of the experimental setup showing fiber launch (FL), photodiodes (PD's) and piezo mirrors (M's).}
    \label{fig:setup_b}
    \end{figure}

Figures \ref{fig:setup_a} and \ref{fig:setup_b} illustrate the experimental setup. A 1550 nm (C band) laser feeds light into a circulator, and the light then exits via a fiber launch (FL) and enters the first beamsplitter (BS1). After passing through the Butler matrix, four distinct beams eventually retroflect from their corresponding mirrors (M1, M2, etc.), which are piezo driven and used to modulate the phase of each channel. Consequently, we consider this the input of the Butler Matrix, with the initial pass through the GM simply functioning as a convenient method for splitting light into four channels and providing alignment beams for back-reflection. The modulated beams return through the network, interfering as expected. Photodetectors are placed at the outputs to measure the resulting interference pattern. Channel 1 returns through the fiber launch and is rerouted through the circulator to another photodetector.\par

We used non-polarizing beamsplitter cubes manufactured by Lambda Research Optics (Costa Mesa, CA) and anti-reflection coated for operation at 1550 nm. Each cube was specified by the manufacturer with a T and R value of $50\% \pm3 \%$ \cite{noauthor_broadband_nodate}. After alignment, we achieved a fringe visibility between 95\% and 99\% for all $\binom{4}{2}=6$ binary interferometric combinations.\par

We first attempted solving our system with an interior point algorithm \cite{boyd_convex_2004}. As noted above, however, small amounts of hysteresis in the phase modulation and temperature gradients across the experimental apparatus induce phase noise, which degraded the algorithm's performance. By extension, any algorithm relying on numerical gradient approximations will naturally be sub-optimal as well since the error space is constantly changing. \par

To combat the stochastic nature of the device we implemented a constrained, globalized, and bounded Nelder-Mead method (GBNM) \cite{luersen_constrained_2004}. The traditional Nelder-Mead Simplex algorithm is gradient-free and tends to perform better than other descent alogrithms over nonsmooth cost functions \cite{nelder_simplex_1965}. The GBNM variation allows for constraints and restarts if the algorithm spends too many iterations around a particular point. This restart feature is key for overcoming noise-induced local minima. \par

\begin{figure*}
\centering
\includegraphics[width=\figWidth]{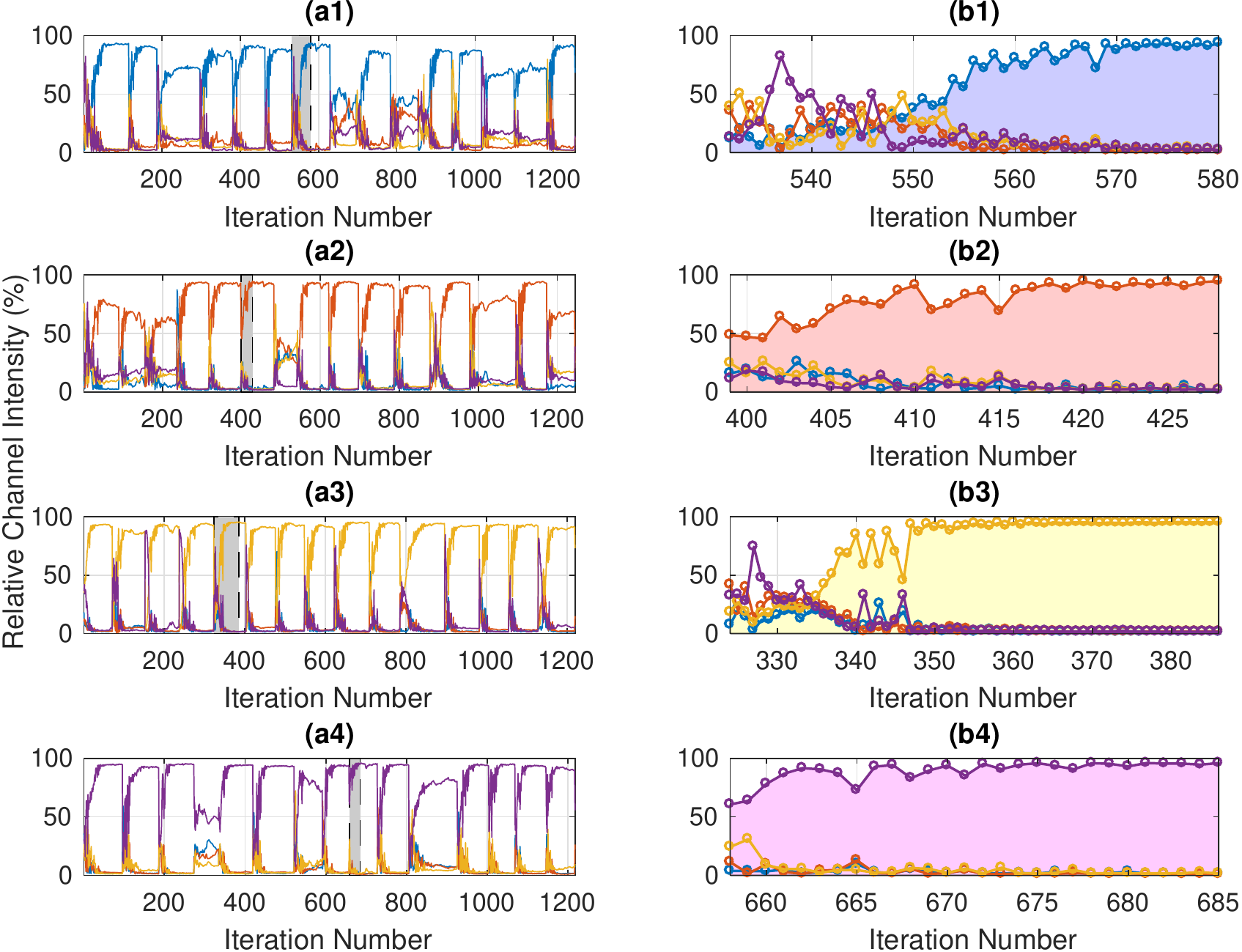}
\caption{Evolution of relative channel intensities while maximizing channels 1 (a1), 2 (a2), 3 (a3), and 4 (a4) using the GBNM algorithm. Before running, the algorithm chose 12 random initialization points inside of the constraint sphere (defined by the piezoelectric voltage driver). It began descending from each point for 100 iterations, after which it moved on to the next point (unless it converged). Certain initializations converged much deeper than others, depending on how close they were to a noise induced local minimum. The relative channel intensities were saved at each iteration. The best initial point, along with its evolution for a particular channel are shown in column (b). Final channel intensities were 93.7\% (b1), 94.7\% (b2), 95.4\% (b3), and 96.0\% (b4).}
 \label{fig:data}
\end{figure*}

Figure \ref{fig:data} details the evolution of the GBNM algorithm for each channel. Optimal values for channels 1, 2, 3, and 4 result in 93.7\%, 94.7\%, 95.4\% and 96.0\% relative intensity respectively. The restart nature of the GBNM algorithm is evident by the repeated jumps from high to low channel intensity. In this particular configuration, 12 points were randomly chosen at initialization. If the algorithm didn't converge after 100 iterations, it moved onto the next starting point. \par

Given the imperfect beamplitting ratios and fringe visibility, error analysis indicates that deeper convergence (i.e channel intensities of 100\%) requires beamsplitters with tolerances closer to the ideal 50\% splitting ratio to achieve better fringe visibility. Despite these imperfections, the method appears to accurately characterize the codebook of the device.

\section{Conclusion and Future Work}
Designing and operating Green Machines is difficult because of stringent phase matching requirements. We derived and demonstrated a simple method to compensate for phase inconsistencies and characterize a device after fabrication using a feedback algorithm.\par 

Our feedback algorithm scales well with size and does not require previous knowledge of the system architecture, since it relies on a convex objective function as defined by the inherent nature of the GM. Several optimization routines are readily available to minimize the system's objective function. We demonstrated the practicality of using the GBNM algorithm to compensate for time-varying system fluctuations and avoid noise-induce local minima. Future work may explore other routines that better handle these system fluctuations.\par

While our current configuration is in free space, future work will implement this algorithm on an integrated device and explore scaling with the size of the Butler matrix.

%
%
%

\section*{Acknowledgements}
We thank Karl Warnick for useful discussions related to this work and in particular pointing us to previous studies of RF Butler matrices.

\bibliographystyle{unsrt}
\bibliography{zotero_v2}

%
%

\end{document}